\documentclass[conference]{IEEEtran}
\IEEEoverridecommandlockouts
\usepackage{cite}
\usepackage{amsmath,amssymb,amsfonts}
\usepackage{algorithmic}
\usepackage{graphicx}
\usepackage{textcomp}
\usepackage{xcolor}
\def\BibTeX{{\rm B\kern-.05em{\sc i\kern-.025em b}\kern-.08em
    T\kern-.1667em\lower.7ex\hbox{E}\kern-.125emX}}

\usepackage[T1]{fontenc}
\usepackage[utf8]{inputenc} % arXivでも大丈夫
\usepackage{lmodern}        % Latin Modern フォントにする

\begin{document}

\title{Long-Term Variability in Physiological-Arousal Relationships for Robust Emotion Estimation\\
%{\footnotesize \textsuperscript{*}Note: Sub-titles are not captured in Xplore and should not be used} % so delete this line
}

\author{\IEEEauthorblockN{Hiroto Sakimura}
\IEEEauthorblockA{\textit{Bioinspired Systems Research-Domain}\\
\textit{Toyota Central R\&D Labs., Inc.}\\
Aichi, Japan \\
e1774@mosk.tytlabs.co.jp}
\and
\IEEEauthorblockN{Takayuki Nagaya}
\IEEEauthorblockA{\textit{Bioinspired Systems Research-Domain}\\
\textit{Toyota Central R\&D Labs., Inc.}\\
Aichi, Japan \\
nagaya@mosk.tytlabs.co.jp}
\and
\IEEEauthorblockN{Tomoki Nishi}
\IEEEauthorblockA{\textit{Social Systems Research-domain}\\
\textit{Toyota Central R\&D Labs., Inc.}\\
Aichi, Japan \\
nishi@mosk.tytlabs.co.jp}
\and
\IEEEauthorblockN{Tetsuo Kurahashi}
\IEEEauthorblockA{\textit{Bioinspired Systems Research-Domain}\\
\textit{Toyota Central R\&D Labs., Inc.}\\
Aichi, Japan \\
kurahashi@mosk.tytlabs.co.jp}
\and
\IEEEauthorblockN{Katsunori Kohda}
\IEEEauthorblockA{\textit{Bioinspired Systems Research-Domain}\\
\textit{Toyota Central R\&D Labs., Inc.}\\
Aichi, Japan \\
e1100@mosk.tytlabs.co.jp}
\and
\IEEEauthorblockN{Nobuhiko Muramoto}
\IEEEauthorblockA{\textit{Bioinspired Systems Research-Domain}\\
\textit{Toyota Central R\&D Labs., Inc.}\\
Aichi, Japan \\
muramoto@mosk.tytlabs.co.jp}
}

% \and
% \and
% \and
% \and
% \and

\maketitle

\begin{abstract}
Estimating emotional states from physiological signals is a central topic in affective computing and psychophysiology. While many emotion estimation systems implicitly assume a stable relationship between physiological features and subjective affect, this assumption has rarely been tested over long timeframes. This study investigates whether such relationships remain consistent across several months within individuals.
We developed a custom measurement system and constructed a longitudinal dataset by collecting physiological signals—including blood volume pulse, electrodermal activity (EDA), skin temperature, and acceleration—along with self-reported emotional states from 24 participants over two three-month periods. Data were collected in naturalistic working environments, allowing analysis of the relationship between physiological features and subjective arousal in everyday contexts.
We examined how physiological–arousal relationship evolve over time by using Explainable Boosting Machines (EBMs) to ensure model interpretability. A model trained on 1st-period data showed a 5\% decrease in accuracy when tested on 2nd-period data, indicating long-term variability in physiological–arousal associations. EBM-based comparisons further revealed that while heart rate remained a relatively stable predictor, minimum EDA exhibited substantial individual-level fluctuations between periods.
While the number of participants is limited, these findings highlight the need to account for temporal variability in physiological–arousal relationships and suggest that emotion estimation models should be periodically updated—e.g., every five months—based on observed shift trends to maintain robust performance over time. \footnote{This manuscript has been accepted at 13th International Conference on Affective Computing and Intelligent Interaction (ACII 2025).}
\end{abstract}


\section{Introduction}

Emotion estimation systems are one of the central topics in affective computing, human–computer interaction, and real-world mental health monitoring. Among various modalities—including facial expressions, vocal cues, and behavioral patterns—physiological signals such as heart rate variability (HRV), electrodermal activity (EDA), and skin temperature have attracted growing attention due to their objectivity and robustness across contexts \cite{picard2000affective, can2023approaches, shukla2019feature, appelhans2006heart, egger2019emotion}. Recent advances in wearable devices have further enhanced the feasibility of physiological sensing by enabling non-invasive, continuous monitoring without relying on external environmental conditions such as lighting or noise \cite{healey2005detecting, ragot2018emotion, borrego2019reliability, chandra2021comparative, zhao2018emotionsense, saganowski2022emotion, guo2015short}. These advantages make physiological signals particularly promising for emotion recognition in naturalistic environments, including workplaces. 

Despite substantial progress in physiological-based emotion recognition, most existing studies have focused on short-term experiments in controlled laboratory or online settings \cite{DVN/R9WAF4_2021, sharma2019dataset, gatti2018emotional, behnke2022autonomic}, offering limited insight into the temporal dynamics of physiological-emotional relationships in real-world environments \cite{kim2008emotion, can2023approaches, bota2024real, d2023affect}. Although several researchers have acknowledged the potential influence of contextual and temporal factors—such as daily routines, seasons, or stress levels—few have provided empirical evaluations of how such variations affect model performance over time \cite{beatton2024positive, borges2020emotion, quigley2014there}. 
Moreover, recent research has highlighted that subject-dependent models outperform subject-independent ones in affective state prediction, particularly when using multimodal and longitudinal data collected in daily-life contexts \cite{berkemeier2025}. These findings support the growing consensus that individual-level modeling enhances prediction accuracy by accounting for inter-individual variability. However, such models often implicitly assume that the physiological-affective relationship within individuals remains stable over time. 

To address this gap, the present study investigates whether the relationship between physiological signals and subjective emotion remains consistent over time within individuals. This study focuses on arousal, which is thought to have association with autonomic nervous activity inferred from physiological indices \cite{moreira2018emotional, bradley2008pupil}. We constructed a longitudinal dataset by collecting physiological and self-reported emotional data from 24 participants in real-world working environments over two separate three-month periods. Physiological signals—including blood volume pulse (BVP), EDA, skin temperature, and acceleration—were continuously measured using wearable sensors, while subjective emotion was recorded using a custom smartwatch application based on the Core Affect Model \cite{russell1999core}. To analyze temporal shifts in physiological-emotional associations, we employed an Explainable Boosting Machine (EBM), a model capable of capturing both global trends and time-specific variations. The results revealed a 5\% drop in model accuracy when training on the 1st period and testing on the 2nd, highlighting temporal instability—that is, long-term variability in physiological–arousal associations. Furthermore, we evaluated the consistency of each physiological feature in predicting arousal and proposed potential adaptation strategies to maintain model performance over time.
%Notably, heart rate (HR) was the most consistent predictor of arousal, while minimum EDA exhibited the greatest variability. 
%These findings suggest that emotion estimation models should be updated at least every five months to maintain reliability in real-world applications.

\section{Related Works}

\subsection{Physiological Signals and Emotion Estimation}
Physiological signals have long been studied as reliable indicators of emotional states in both psychophysiology and affective computing. HRV, EDA, and skin temperature have been widely utilized due to their direct connection to the autonomic nervous system \cite{cacioppo1990inferring, picard2000affective, moreira2018emotional, bradley2008pupil,healey2005detecting, ragot2018emotion, borrego2019reliability, chandra2021comparative, zhao2018emotionsense, saganowski2022emotion, guo2015short, can2023approaches, shukla2019feature, appelhans2006heart, egger2019emotion}. HRV reflects the balance between sympathetic and parasympathetic activity and has been shown to correlate with arousal and emotional regulation \cite{kim2008emotion, zhao2018emotionsense, beatton2024positive}. EDA, which is driven by sympathetic activity, is especially sensitive to changes in arousal and stress levels \cite{healey2005detecting, bradley2008pupil, shukla2019feature}. These modalities have proven effective in lab-based emotion recognition settings when combined with machine learning models \cite{picard2001toward, can2023approaches}.

In addition, multimodal datasets combining physiological and behavioral signals have facilitated the development of emotion-aware systems in increasingly complex contexts \cite{siddiqui2022survey, sharma2019dataset}. Despite technological advancements, most studies still rely on data collected under controlled conditions or short recording windows \cite{DVN/R9WAF4_2021, sharma2019dataset, gatti2018emotional, behnke2022autonomic}.

\subsection{Temporal Variability in Physiological-Emotional Associations}
Emerging research has begun to question the temporal stability of the relationship between physiological states and emotions. Recent findings suggest that emotion-related physiological patterns may vary with daily activities, seasons, and stress regulation capacity \cite{beatton2024positive, borges2020emotion, weber2010low}. Circadian and seasonal rhythms also influence autonomic activity, complicating the assumption of a fixed mapping between physiology and affect \cite{picard2000affective, can2023approaches}.

Several studies have explored emotion estimation ``in the field'' %\sout{"in the wild"} 
using wearable devices, mobile applications and games \cite{healey2010out, zenonos2016healthyoffice, exler2016wearable, d2023affect, kutt2022biraffe2}. However, these efforts typically span only a few days to four weeks, limiting the ability to assess within-subject variability over longer periods.

\subsection{Limitations of Existing Emotion Estimation Models}
Many existing models assume a static, person-independent relationship between physiological signals and emotional states. These general-purpose models often perform well under controlled experimental conditions but suffer from performance degradation in real-world, long-term settings \cite{can2023approaches, d2018affective, d2023affect}. This limitation is partly due to aggregated datasets that dilute individual differences and overlook within-subject fluctuations over time.

Although some studies have proposed adaptive modeling approaches—such as incremental learning or transfer learning—few have examined how feature–emotion associations shift over extended timeframes \cite{quigley2014there}. In particular, to our knowledge, no prior work has quantitatively assessed intra-individual temporal variability using repeated measurements over multiple months, although \cite{berkemeier2025} performed a long-term data collection spanning nine months.

\section{Materials and Methods}

\subsection{Participants and Data Collection Duration}
Twenty-four Japanese-speaking participants were recruited for this study, evenly distributed across four age groups: 25–34 (Participant ID: A-F), 35–44 (G-L), 45–54 (M-R), and 55–64 (S-X) years. Each age group included three males and three females, resulting in a balanced sample by both age and gender. All participants were full-time employees working in office environments. Data collection was conducted during two distinct three-month periods: the 1st period from November 1, 2023 to February 9, 2024, and the 2nd period from April 15 to July 18, 2024. During each period, physiological and subjective data were collected every weekday throughout participants' regular working hours, approximately 8:30 to 17:30, totaling eight hours per day. This longitudinal design enabled within-subject comparison of physiological-emotional relationships across different periods. The study protocol was approved by the ethics committee of Toyota Central R\&D Labs., Inc.

\subsubsection{Subjective Emotion Annotation}
Subjective emotional states were annotated using a custom-developed application installed on an Apple Watch. To ensure ecological validity and minimize recall bias, the system was designed to prompt emotion input in real-time during natural working conditions. Specifically, when the accelerometer detected a stationary state lasting for 50 seconds, the device provided a gentle vibration cue, prompting the participant to report their emotional state immediately via the watch interface.

Data entries were automatically triggered based on this detection criterion, ensuring temporal synchronization between physiological signals and self-reported emotions. This approach enabled direct pairing of emotional annotations with preceding 50-second windows of physiological data (which are expected to be noiseless), forming a structured dataset suitable for supervised learning. The frequency of such stationary detections varied across participants, ranging from approximately 5 to 40 times per day.

The annotation interface was grounded in the Core Affect Model proposed by Russell and Barrett \cite{russell1999core}, which defines emotions along two primary dimensions: valence (pleasant–unpleasant) and arousal (activation–deactivation) as shown in Fig. \ref{fig1}(a). The input procedure followed a two-step structure. In the first step, participants selected one of four general emotion categories (Happy, Nervous, Sad, or Relaxed) that represent a quadrant in the affective space. In the second step, they chose one of four specific emotion labels corresponding to the selected category. All emotion options were displayed in Japanese to ensure intuitive comprehension by Japanese speakers.

Importantly, emotion intensity was not recorded in the main study. A preliminary test with a four-level intensity scale revealed low variance in ratings during typical working conditions. To reduce participant burden and streamline interaction, the intensity input was excluded from the final protocol.

\subsubsection{Physiological Data Collection}
\begin{figure}
\centering
\includegraphics[width=9cm]{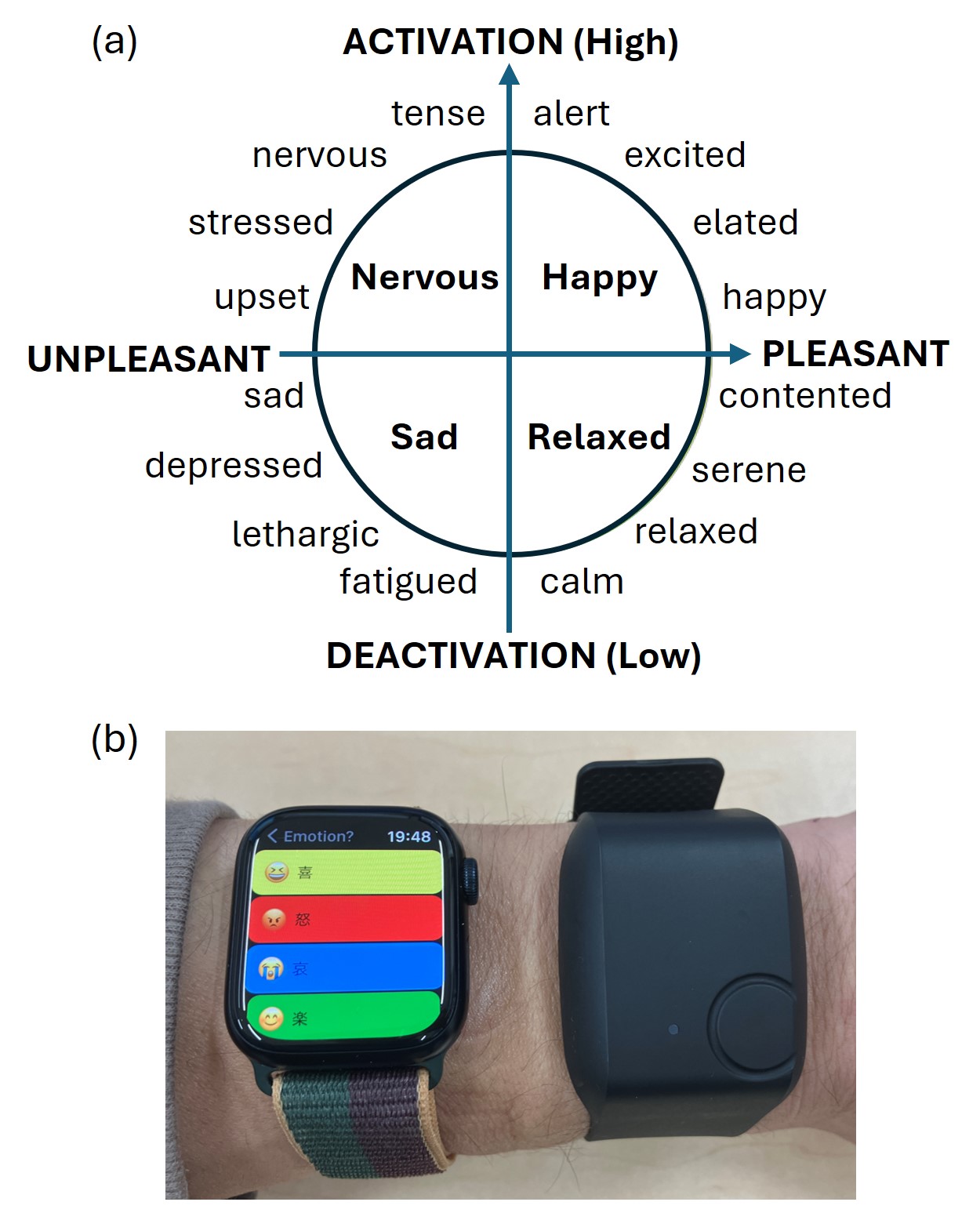}
\caption{(a) The core affect model as selections of subjective emotion \cite{russell1999core}. (b) The state in which two devices are worn. Left: Apple Watch to record subjective emotions. Right: Empatica E4 Wristband for physiological measurement. }
\label{fig1}
\end{figure}

Physiological data were continuously recorded throughout each working day using the Empatica E4 wristband, a multi-sensor wearable device widely used in affective computing research \cite{ragot2018emotion, borrego2019reliability, chandra2021comparative, zhao2018emotionsense}. The following physiological signals were collected:
\begin{itemize}
    \item Photoplethysmogram (PPG, 64 Hz): Measures BVP to derive HRV, which is an indicator of autonomic nervous system activity.
    \item Electrodermal Activity (EDA, 4 Hz): Measures skin conductance levels, reflecting sympathetic nervous system activity.
    \item Skin Temperature (4 Hz): Captures the infrared radiation emitted from the skin's surface, allowing it to determine the skin's temperature.
    \item Acceleration (32 Hz): Measures tri-axial wrist movement %\sout{movement of the wrist-worn device} 
    to assess physical activity of the E4-equipped arm.
\end{itemize}
All signals were recorded at sampling rates noted above and stored locally before being uploaded for processing. Measurements were conducted continuously during working hours across both data collection periods. The participants wore two devices as shown in Fig. \ref{fig1}(b). %To ensure data quality, artifacts due to motion or signal loss were subsequently removed during preprocessing (described in Section \ref{sec:pre}).

\subsection{Preprocessing and Feature Extraction}
\label{sec:pre}
To prepare the physiological data for modeling, a series of preprocessing and feature extraction steps were applied to the 50-second segments preceding each emotion annotation. These segments were identified based on stationary periods detected via accelerometry, ensuring minimal movement artifacts.

Preprocessing: BVP signals occasionally exhibited noise artifacts even during stationary periods. Segments with abnormal signal patterns were visually inspected and manually excluded. A low-pass Butterworth filter with a 3 Hz cutoff was applied to the BVP data to remove high-frequency noise. Features were calculated from the inter beat intervals (IBI) derived from BVP. Preliminary inspection of the other signals—EDA, skin temperature, and acceleration—indicated negligible high-frequency noise during stationary conditions; therefore, no additional filtering was applied to those channels.

Segmentation and Feature Extraction: For BVP, each 50-second segment was divided into overlapping 30-second windows with a 5-second stride. For each window, a set of statistical and domain-specific features was computed, and the average across all windows was used to represent the segment. For other three measures, we treated the original 50-second segment as one window and extracted features. Features of acceleration were extracted from the data 240 to 50 seconds before the emotion was recorded, when participants were expected not to be in the stationary state. In total, 17 features were extracted across four signal modalities: cardiovascular (e.g., RMSSD, pNN50, HR), electrodermal (e.g., maximum and mean EDA), thermal (mean skin temperature), and motion-related (mean and maximum acceleration). A full list of extracted features is provided in Table \ref{tab:features}. The extracted features were chosen based on their physiological interpretability and prior relevance to arousal in affective computing research (e.g., HRV metrics, EDA, temperature, and acceleration) \cite{appelhans2006heart, zhao2018emotionsense, bradley2008pupil, zenonos2016healthyoffice, kreibig2010autonomic}. Before proceeding to the next step, outliers were removed for each participant and feature using a threshold of 3$\sigma (\sigma$ is standard deviation).

\begin{table}[t]
    \centering
    \caption{Extracted features from each physiological signals: BVP, EDA, skin temperature and acceleration. }
    \begin{tabular}{|c|c|}  \hline
         Feature Name & Description  \\ \hline \hline
        
         SD & Standard deviation of IBI  \\ \hline
         CV & SD normalized by mean IBI  \\ \hline
         RMSSD & Root mean square of successive differences of IBI  \\ \hline
         pNN50 & \begin{tabular}{c}
                    Percentage of successive IBIs \\ 
                    differing by more than 50 ms
                \end{tabular}  \\ \hline
         HR & Heart rate (beats per minute)  \\ \hline
         L & Long axis of Poincaré plot \cite{toichi1997new} \\ \hline
         T & Short axis of Poincaré plot \cite{toichi1997new} \\ \hline
         LF & \begin{tabular}{c}
                    Power of low frequency band (0.04–0.15 Hz) \\ 
                    of IBI spectrum
                \end{tabular}  \\ \hline
         HF & \begin{tabular}{c}
                    Power of high frequency band (0.15–0.4 Hz) \\
                    of IBI spectrum
                \end{tabular}  \\ \hline 
         LF/HF & Ratio of LF to HF components  \\ \hline
        
         EDA\_ave & Mean EDA  \\ \hline
         EDA\_max & Maximum EDA  \\ \hline
         EDA\_min & Minimum EDA \\ \hline
         EDA\_diff & Difference between max and min EDA  \\ \hline
        
         Temp\_ave & Mean skin temperature  \\ \hline
        
         Acc\_ave & Mean acceleration magnitude    \\ \hline
         Acc\_max & Maximum acceleration magnitude  \\ \hline
         
    \end{tabular}
    \label{tab:features}
\end{table}

\subsection{Feature Selection}
\label{sec:feature}
We adopted sequential feature selection (SFS) to ensure consistency and interpretability, which is well-suited for our use of EBM. While model-dependent, SFS offered a practical balance between accuracy and transparency and is commonly used in physiological emotion recognition tasks \cite{pedregosa2011scikit, jenke2014feature}. %\sout{To reduce feature dimensionality and enhance model interpretability, we employed sequential feature selection (SFS), a wrapper-based technique commonly used in physiological emotion recognition tasks \cite{pedregosa2011scikit, jenke2014feature}.} 
Feature selection was conducted on the combined dataset across all participants and both data collection periods, rather than on a per-individual or per-period basis. This global approach aimed to identify features with consistent predictive power across different conditions.

The target variable for feature selection was binary arousal, derived from the vertical axis of the Core Affect Model \cite{russell1999core}. Each emotion label was mapped to either a high or low arousal category based on its position in the affective space. Using this binary classification target, SFS was performed to iteratively select a subset of features that optimized classification performance. The resulting five features—HR, Temp\_ave, Acc\_ave, EDA\_min, and EDA\_max—were used consistently in all subsequent modeling and analysis procedures.

\subsection{Modeling and Analysis}

To investigate the temporal variability in the relationship between physiological signals and subjective arousal, we adopted the Explainable Boosting Machine (EBM) \cite{nori2019interpretml}. EBM integrates the transparency of Generalized Additive Models (GAMs) with the flexibility of gradient-boosted decision trees, making it particularly suitable for applications where both interpretability and non-linearity are critical. In our context, understanding how each physiological feature contributes to arousal—potentially in a non-linear manner—and how these relationships change over time is essential.

The EBM model estimates the log-odds of the arousal state as a sum of additive contributions from individual features and their temporal interactions. The model takes the following form:
\begin{align}
    g(E(y)) = \sum_{i=1}^5 f_{com_i}(X_i) + \delta_p \sum_{i=1}^5 f_{int_i}(X_i)
\label{eq1}
\end{align}
where:
\begin{itemize}
    \item $y$ is the binary arousal label (1: high, 0: low),
    \item $g$ is the link function (logit),
    \item $X_i$ denotes the $i$-th physiological features selected via SFS (HR, Temp\_ave, Acc\_ave, EDA\_min and EDA\_max),
    \item $f_{com_i}$ represents the period-independent contribution of feature $X_i$,
    \item $f_{int_i}$ captures the specific temporal shift in the 2nd period,
    \item $\delta_p$ is a dummy variable indicating the period (0: 1st, 1: 2nd).
\end{itemize}

This formulation allows us to disentangle general feature effects from those that are specific to the 2nd period (if there is no shift between two periods, $f_{int_i}(X_i)$ becomes zero), thereby capturing temporal shifts in feature–arousal relationships.

For each participant, we randomly sampled 90 data from both the 1st and 2nd periods and trained the model 100 times to account for sampling variability. Model performance was evaluated using the classification accuracy and area under the receiver operating characteristic curve (AUC). All test data was separated from training data (90 per period) to validate the model performance. To identify significant temporal changes, we examined the 95\% confidence intervals (CIs) of the additive functions $f_{com_i} + f_{int_i}$. Non-overlapping intervals between $f_{com_i}$ and $f_{com_i} + f_{int_i}$ %\sout{periods} 
were interpreted as meaningful shifts in the feature’s contribution to arousal prediction, which provides an intuitive interpretation of the results.

To evaluate how the relationship between physiological features and subjective arousal varies over time more quantitatively, we introduced Pearson’s correlation coefficient $r$ as an additional measure of temporal shift. Specifically, for each feature and participant, we computed the correlation between the period-invariant additive function $f_{com_i}$ and the total contribution during the 2nd period $f_{com_i}+f_{int_i}$, both estimated by the EBM. The Pearson's correlation coefficient $r$, with values closer to 1 indicating stability and lower values suggesting change,  serves as a proxy for functional similarity over time between these additive functions. It enables interpretable assessment of temporal shift in feature–arousal relationships.
%\sout{This coefficient reflects the degree to which the shape of the feature–arousal relationship is preserved across periods}
This coefficient is sensitive to, for example, the shift in the $x$ direction, but insensitive to the scaling or shifting to the $y$ direction. %This use of $r$ as a functional stability metric represents a novel contribution of the present study. 

\section{Results}

\subsection{Dataset Overview and Participant Selection}
\begin{figure}
\centering
\includegraphics[width=7cm]{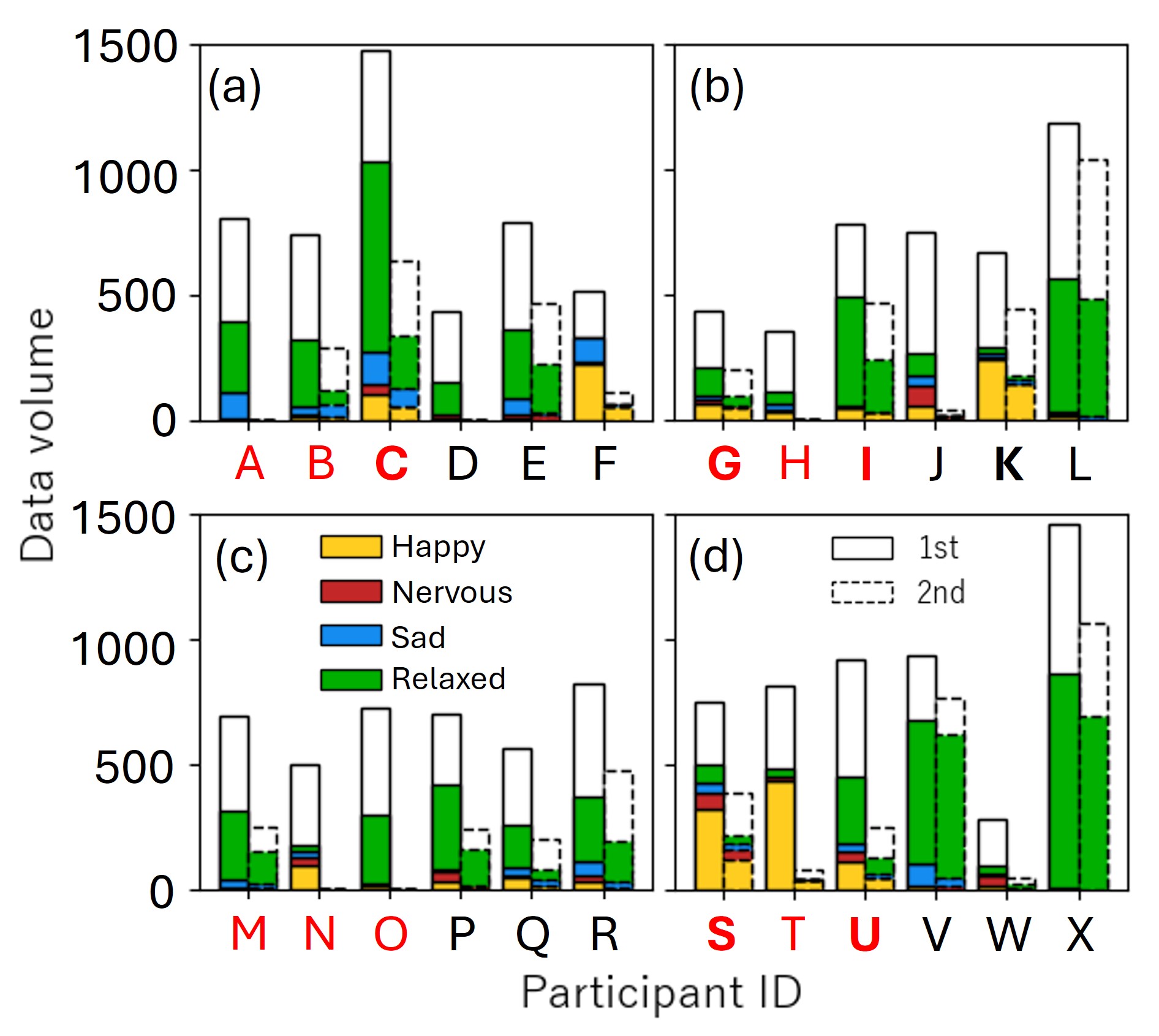}
\caption{Data amount for each participant in the 1st and 2nd period divided by the participants' age group; (a) 25-34, (b) 35-44, (c) 45-54 and (d) 55-64. Bars are color-coded by emotion category. White bars represent the data eliminated due to noise mixture}.  Red ID indicate female participants. %\sout{The colored bars represent the emotional breakdown that was not eliminated in the preprocessing process.} 
The bold six participants satisfied the requirements for additional analysis.
\label{fig2}
\end{figure}

Figure \ref{fig2} summarizes the total number of usable emotion-labeled segments for each participant, along with their emotional distributions across the two study periods. Almost half of the data were eliminated due to noise mixture as indicated by empty bars in the figure. The majority of the collected data corresponds to low-arousal and positive-valence states, particularly the ``Relaxed'' category, reflecting the natural affective profile during typical working conditions. While the emotional label usage varied across individuals—some favoring Happy, others Relaxed—no consistent trends were found based on age or gender. This suggests that emotional reporting was more influenced by personal context than demographic factors. Five participants did not participate in the 2nd data collection period, and their data was used only for feature selection process. 

%Based on the results of feature selection (Section \ref{sec:feature}), five physiological features—heart rate (HR), mean skin temperature (temp ave), mean acceleration (acc\_ave), minimum EDA (EDA\_min), and maximum EDA (EDA\_max)—were selected as the most informative predictors of arousal. 
For more detailed within-subject analysis, we focused on six participants (IDs C, G, I, K, S, and U) who exhibited a sufficient number of high and low arousal samples in both periods. Their data were used for all subsequent EBM-based modeling and comparison. Although the six include five females and one male, the absolute number is too small that we do not divide the following discussions by gender or age.

\subsection{Prediction Performance of One-Period Models}
Table \ref{tab} presents the average classification performance of the EBM models under three different train–test configurations for the six participants per each. %\sout{All the test data was separated from the training data.} 
In case (a), the model was trained and tested using data from the 1st period only, achieving a baseline accuracy of 0.771 and an AUC of 0.586. In case (b), the model trained on the 1st-period data was tested on 2nd-period data, resulting in a performance drop to 0.716 in accuracy and 0.548 in AUC. This decline indicates that the relationship between physiological features and arousal was not fully preserved across the two periods.
Case (c) describes the performance of the model trained and tested on the 2nd-period data only, which yielded an average accuracy of 0.731 and AUC of 0.568. These scores exceeds that of the 1st-trained/2nd-tested model, indicating that the observed degradation is attributable to temporal shift rather than intrinsic difficulty of the 2nd-period data.
In case (d), a model trained on the combined data from both periods yielded improved performance on 2nd-period data (accuracy: 0.734, AUC: 0.571) compared to case (b) and (c). This suggests that the model was able to incorporate patterns specific to the 2nd period when trained with mixed data, partially recovering performance.

Collectively, these results suggest the presence of temporal shift in the physiological–emotional relationship. Without model updating, performance degrades over time—highlighting the need for periodic retraining to maintain model reliability in real-world deployments.

\begin{table}
    \centering
    \caption{The accuracy, AUC, and their standard errors (in brackets) in several combination of training and test data. Standard errors reflect 100 repeated train/test splits per participant and were averaged across individuals.} 
    
    \begin{tabular}{c|cc|cc}
        Case & Train & Test &  Accuracy &  AUC \\ \hline          
         (a) & 1st  & 1st  &  0.771 (0.00486) & 0.586 (0.00282)  \\
         (b) & 1st  & 2nd  &  0.716 (0.00598) & 0.548 (0.00377)  \\
         (c) & 2nd  & 2nd  &  0.731 (0.00382) & 0.568 (0.00263)  \\
         (d) & 1st+2nd & 2nd  &  0.734 (0.00534) & 0.571 (0.00374)  \\        
    \end{tabular}   
    \label{tab}
\end{table}

\subsection{Variations over time on Physiological-Emotional Relation}

\begin{figure}
\centering
\includegraphics[width=8cm]{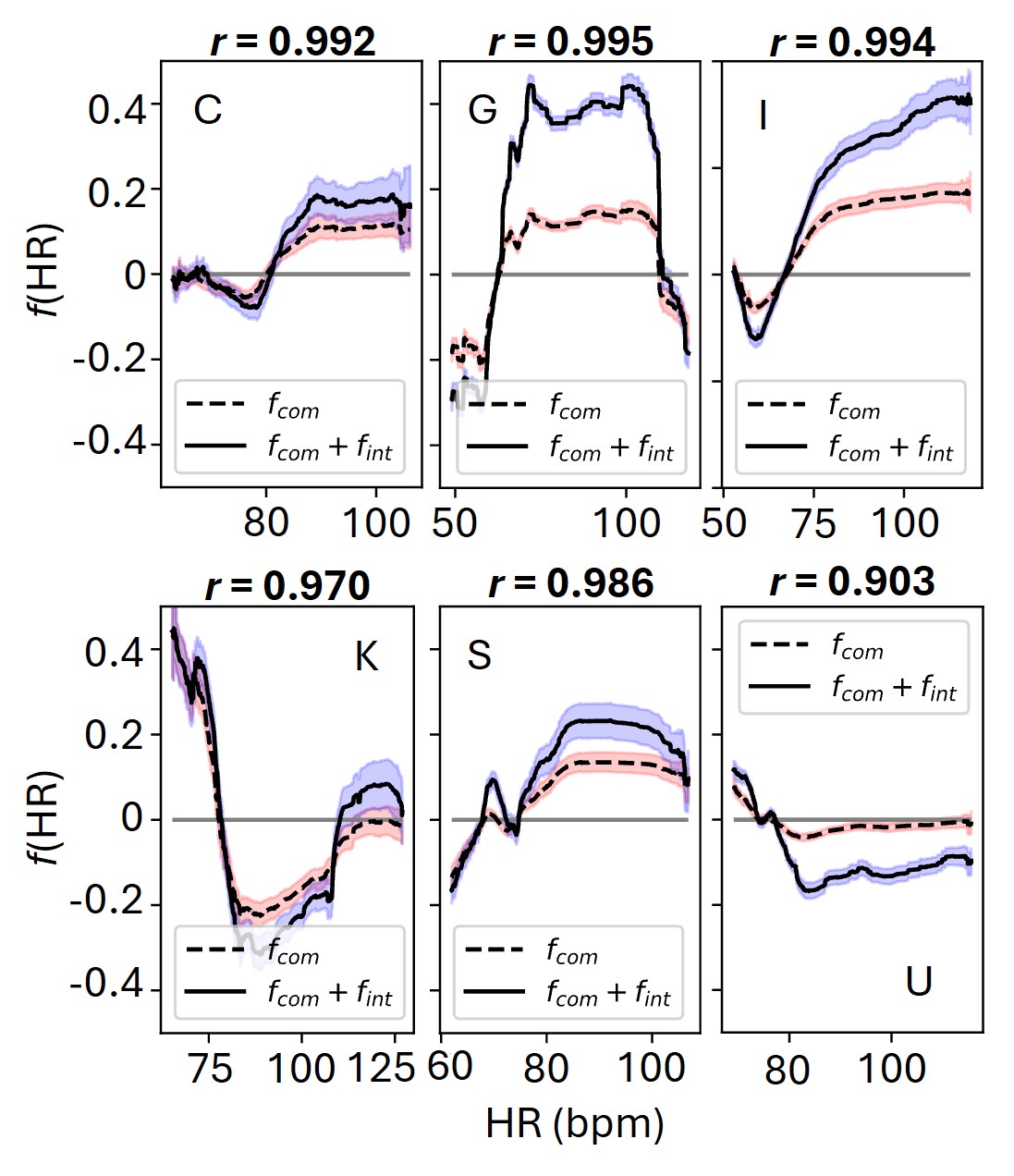}
\caption{Relations between HR and its contribution to predict arousal for the six participants. Alphabets correspond to the participant ID. $f_{com}$ reflects the common contribution of HR to predict arousal regardless of the period, and $f_{com}+f_{int}$ includes the contribution that appears only in the 2nd period. Colored region represents the 95\% confidence interval.}
\label{fig3}
\end{figure}

Fig. \ref{fig3} illustrates the additive functions of HR in predicting arousal for the six participants derived from the fitting of Eq.\ref{eq1}. The visual comparison between 
$f_{com}$ and $f_{com}+f_{int}$, together with their corresponding 95\% CI and $r$ values, reveals participant-specific differences in temporal stability. For example, Participant G exhibited an almost identical functional shape with a high $r$, suggesting only amplitude scaling. In contrast, Participant U showed substantial changes in the shape of the function, reflected by the lowest $r$, indicating a shift in the underlying HR–arousal relationship. These findings demonstrate that the coefficient $r$ aligns well with qualitative observations and is thus suitable for quantifying temporal variation.

\begin{figure}
\centering
\includegraphics[width=8cm]{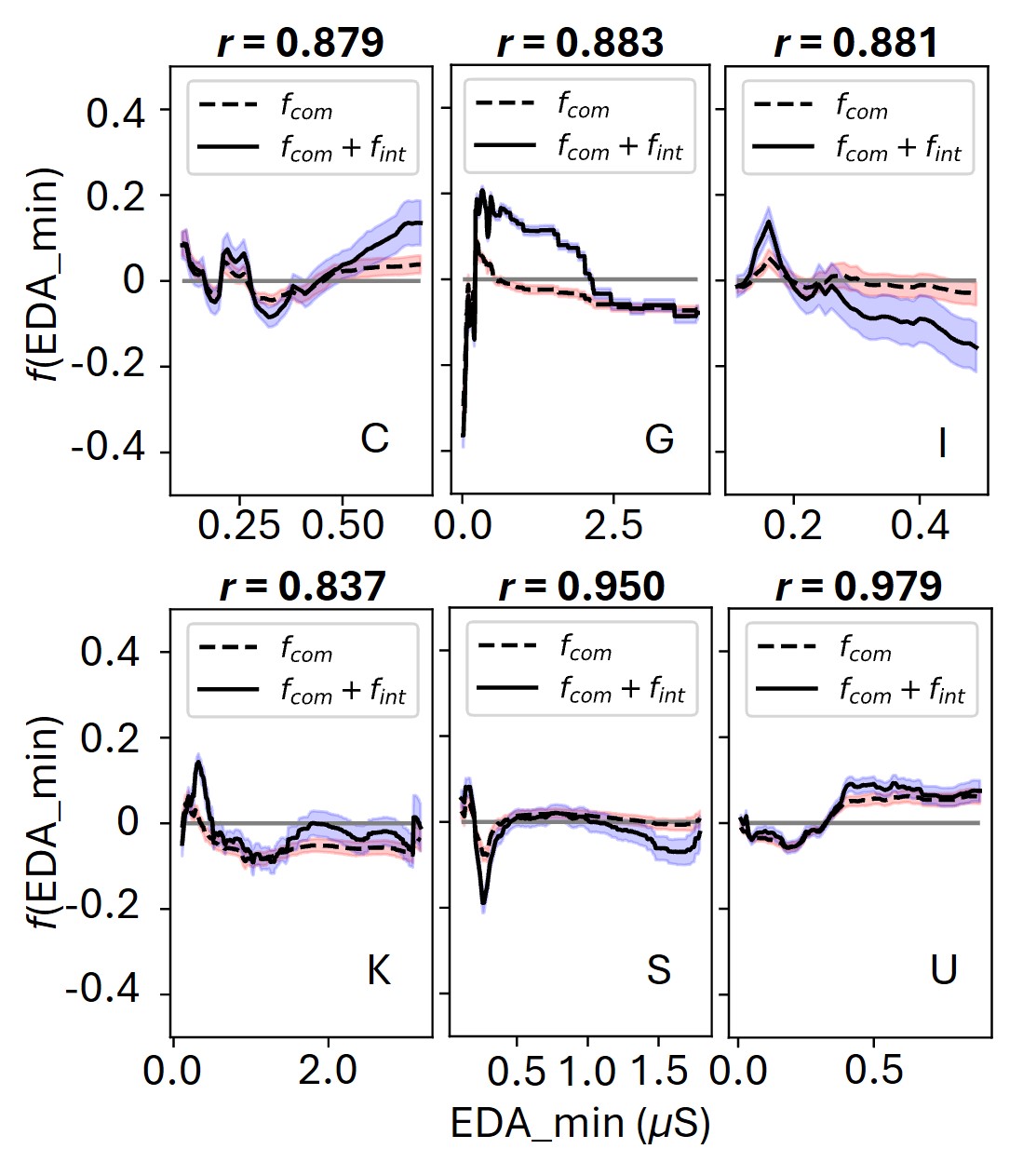}
\caption{Relations between EDA\_min and its contribution to predict arousal for the six participants. Alphabets correspond to the participant ID. $f_{com}$ reflects the common contribution of HR to predict arousal regardless of the period, and $f_{com}+f_{int}$ includes the contribution that appears only in the 2nd period. Colored region represents the 95\% confidence interval.}
\label{fig4}
\end{figure}

Fig. \ref{fig4} presents a similar analysis for minimum EDA (EDA\_min), a feature that exhibited greater variability across periods. Participants G and I, for instance, showed a notable increase in the predictive contribution of EDA\_min during the 2nd period, shifting from near-zero influence to strong positive association. The corresponding drop in $r$ values reflects these changes, supporting the validity of $r$ as a measure of temporal shift in feature–arousal associations.

\begin{figure}
\centering
\includegraphics[width=6cm]{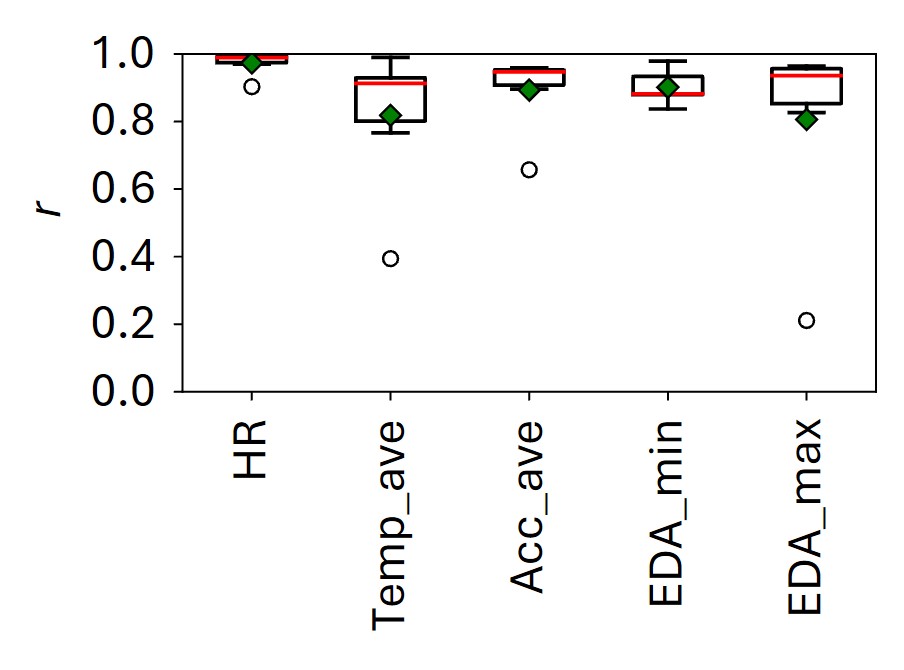}
\caption{Cross-period variability characterized by correlation coefficient $r$ of all selected features. Red lines and green diamonds represent the median and mean values of each feature, respectively.}
\label{fig5}
\end{figure}

To compare temporal stability among all selected features, we aggregated the $r$ values across participants. Fig. \ref{fig5} shows the result plotted as box plots with the median drawn in the red line. HR exhibited the highest median, indicating consistent predictive utility over time, while EDA\_min showed the lowest median $r$, reflecting considerable variability. Here, we used median to compare feature stability because mean values plotted by green triangles are affected by outliers due to a small sample number. Temp\_ave also shows relatively low consistency, or high variability. This analysis confirms that physiological features differ in their temporal robustness and further validates the use of the correlation coefficient as a feature-level shift indicator.

\section{Discussion}
This study investigated the temporal stability of physiological–arousal relationships and evaluated how time-dependent variability affects the performance of emotion estimation models. The key findings include: (1) models trained on one period performed significantly worse when tested on data from another period, suggesting temporal shift in feature–arousal relationships that had been confirmed by the shift of additive functions in the EBM; (2) the use of Pearson’s correlation coefficient $r$ between the additive functions of the EBM effectively quantified the degree of temporal shift in individual features; and (3) among the selected features, HR remained the most stable predictor, while EDA\_min exhibited the largest variability.

Among the five selected features, EDA\_min exhibited the highest degree of temporal variability. This observation may reflect the physiological and environmental sensitivity of EDA. EDA is known to be modulated by skin conductance, which itself is affected by thermoregulatory sweating, ambient temperature, humidity, and the hydration state of the skin. These factors are influenced by seasonal conditions such as air temperature and clothing, as well as individual conditions such as physical fatigue, stress, and sleep quality. For instance, in warmer months, baseline EDA may rise due to increased perspiration and higher skin conductance, while winter may lead to suppressed activity due to vasoconstriction and lower ambient humidity. The pronounced cross-period shift observed in EDA\_min may thus be partially attributed to these external and internal physiological mechanisms.

The degree of temporal variation also differed markedly between the participants. While some individuals showed minimal shift, others exhibited substantial changes in feature–arousal relationships. Although this study focused on within-subject analysis, the emerging patterns suggest that between-subject factors—such as age, gender, lifestyle, or daily stress exposure—may influence the degree of physiological-arousal stability. For example, younger individuals with more variable daily routines or higher mobility may exhibit larger fluctuations. Future studies with larger and more demographically diverse samples should explore whether such group-level trends exist and can inform adaptive model design.

To evaluate temporal shift quantitatively, we introduced Pearson’s correlation coefficient $r$ as a metric that compares the shape of EBM additive functions between two periods. The observed $r$ values corresponded well with visual assessments of function shifts and provided an interpretable, participant-level measure of stability. For example, HR showed consistently high $r$ values across participants, while EDA\_min had low $r$, reflecting its unstable nature. These results not only underscore the importance of periodic model updating but also highlight the potential influence of seasonal and environmental factors. In practice, emotion estimation systems deployed across multiple seasons should account for variations in physical environments such as temperature, humidity, and light exposure, as well as behavioral adaptations like clothing and activity patterns.

Our findings suggest that emotion estimation models may require updates at least every five months to maintain performance. %\sout{However, this 'five month'} 
``Five months'' is based on our experimental criteria including two separated data collection periods. Therefore, rather than adhering to a fixed update cycle, a more practical approach may involve monitoring model performance and initiating updates when degradation is detected. Approximately $5\%$ performance drop was observed in our experiment. Each participant contributed 90 samples per period. With five features, this results in 18 samples per feature, which aligns with successful modeling in the original EBM study \cite{nori2019interpretml} using 569 samples and 30 features ($\sim$19 per feature). Thus, our setup represents a practical lower bound for exploratory modeling. For example, sudden drops in classification accuracy or changes in feature importance may serve as triggers for retraining. This adaptive strategy can help balance the cost of model updating with the need for temporal robustness in real-world applications.

While this study provides the first empirical evidence of long-term variability in physiological–arousal relationships, several limitations should be noted. First, the number of participants included in the within-subject analysis was limited to six, due to data availability constraints. Additionally, although all participants were Japanese speakers, we do not suggest bias in reporting. Rather, we note that emotional patterns and physiological baselines may vary across cultures, affecting generalizability. %\sout{Cultural bias might also be included since all of them were Japanese speakers.} 
Second, contextual factors such as daily physical activity, sleep patterns, and environmental conditions were not explicitly modeled, although they may have contributed to the observed variability. Third, the study was conducted across only two periods (winter and early summer), and it remains unclear whether observed patterns follow circadian, seasonal, or irregular trends.
Fourth, it is important to note that most of the recorded emotional states in this study were relatively weak and occurred during typical working hours. Due to the design of the measurement system—which triggered emotion annotations only after detecting a stationary state of at least 50 seconds—it was difficult to capture moments of intense emotional arousal. This constraint likely biased the dataset toward low-arousal emotional episodes. Future studies should explore improved annotation techniques that enable in-the-moment emotion labeling in more dynamic, everyday contexts.
Furthermore, while EBM was selected for its interpretability, we acknowledge that exploring other models (e.g., random forests or neural networks) and feature selection strategies could provide complementary insights. This remains a valuable direction for future work.
Although valence annotations were collected, we focused on arousal due to stronger physiological associations (e.g., HRV, EDA) \cite{appelhans2006heart, zhao2018emotionsense, kreibig2010autonomic}. %Arousal provided a clearer signal-to-noise ratio for interpreting feature shift. 
Future work will incorporate valence for a more complete affective model.
Incorporating a broader range of contextual data and longitudinal observations are expected in future works to further refine adaptive emotion estimation systems.

\section{Conclusion}

This study examined the long-term variability of physiological–arousal relationships by analyzing how the association between wearable-derived physiological features and subjective emotional records over time. Using a longitudinal dataset collected from naturalistic working environments over two separate three-month periods, we evaluated model performance and feature behavior using EBMs. Our analysis revealed that models trained on data from a single period experienced approximately 5\% drop in accuracy when applied to data from a later period, indicating the presence of temporal shift in physiological–arousal mappings.

Among the five selected features, HR demonstrated the highest temporal stability, whereas EDA\_min showed the greatest variability. To quantify this shift, we introduced Pearson’s correlation coefficient $r$ as a metric to compare the additive functions of each feature over time. This measure effectively captured individual-level variation and allowed for a systematic comparison of feature robustness.

These findings challenge the common assumption of static emotion–physiology relationships in affective computing and highlight the importance of adaptive, personalized models. We recommend that emotion estimation systems be periodically updated—approximately every five months—or adaptively retrained based on performance degradation indicators. Moreover, external factors such as seasonal changes and behavioral context should be considered to improve real-world applicability.

Finally, we emphasize that most of the emotional states recorded in this study were relatively mild and occurred during routine workdays. Due to the stationary-state requirement for physiological data capture, moments of strong emotional arousal may have been underrepresented. Future research should explore annotation techniques that support real-time, in-the-field emotion labeling, and extend the observational period and participant diversity to further validate and generalize these findings.

\section*{Ethical Impact Statement}

This study investigates the temporal variability in physiological–emotional relationships using longitudinal data collected from wearable sensors in real-world work environments.
The participants were recruited internally and took part in the experiment without compensation. They were instructed to wear two wearable devices during their work and to input their emotions at the timings specified by the devices (see main text). Written informed consent was obtained prior to the experiment, including information regarding data handling, withdrawal and other related matters. As stated in the main text, the experiment was approved by the institute's ethics committee (approval number 23A-06); however, no review was conducted by the Institutional Review Board (IRB) as the experiment was non-invasive. As affective computing technologies advance, particularly those involving the detection and modeling of emotional states through physiological signals, it is imperative to consider their ethical implications.

Emotion estimation systems hold promise for applications in mental health support, user-adaptive systems, and workplace well-being monitoring. However, such systems may also be misused for manipulative purposes, such as emotion-based advertising, behavioral nudging, or even surveillance at any place with the help of current information technology. The passive and continuous nature of physiological monitoring further raises concerns about user consent, privacy, and autonomy. Regulatory frameworks and transparency guidelines should be established to govern how emotion-sensitive data are collected, processed, and applied in practice.

Moreover, our dataset is composed exclusively of office workers in Japan, potentially limiting the generalizability of findings across different cultural, occupational, and demographic groups. Emotional expression and physiological responses are known to vary across populations; thus, expanding data collection to more diverse samples is critical for building inclusive and fair emotion-aware systems.

In addition, the study focuses on emotional states that were predominantly mild and captured during sedentary work contexts. Strong or acute emotions were likely underrepresented due to the system’s design, which triggered emotion annotation only during extended stationary states. Future research should explore annotation techniques that allow real-time, in-the-field emotion labeling to capture the full spectrum of affective experiences, especially in dynamic or high-arousal situations.

Finally, as our findings support the need for personalized and temporally adaptive models, it is important to recognize the dual-use potential of such technologies. While personalization can enhance system effectiveness and user alignment, it may also open doors to manipulation or overfitting to individual vulnerabilities. Ensuring ethical use will require continued interdisciplinary dialogue among researchers, ethicists, industry practitioners, and policymakers.

\bibliographystyle{IEEEtran}
\bibliography{citations}

\end{document}